# Lua API and benchmark design using 3n+1 sequences

*Comparing API elegance and raw speed in Redis and YottaDB databases*

Berwyn Hoyt

**Abstract:** Elegance of a database API matters. Frequently, database APIs suit the database designer, rather than the programmer's desire for elegance and efficiency. This article pursues both: firstly, by comparing the Lua APIs for two separate databases, Redis and YottaDB. Secondly, it looks under the API covers at how object orientation can help to retain API efficiency. Finally, it benchmarks both databases using each API to implement a *3n + 1* sequence generator (of Collatz Conjecture fame). It covers the eccentricities of the Lua APIs, the databases, and the nifty choice of benchmark tool, presenting benchmark results of each database's unique design.

## 1 Introduction

This project compares both Redis® and YottaDB® in terms of their respective Lua APIs, assessing API elegance and benchmarking API speed.[1] In comparing the APIs the article will discuss how the use of object orientation contributes to both elegance and efficiency. The benchmark is performed by running a Lua implementation of a *3n + 1* sequence generator (of Collatz Conjecture fame [1]) against both databases. I will cover eccentricities of the Lua APIs, certain significant design choices, the databases, and the nifty choice of benchmark tool.

All of the software showcased in this article is available in public repositories [2].[2] YottaDB and its upstream software, FIS GT.M™, are both open source [3,4]; Redis is available [5], but in 2024 changed from an open source licence to a source-available licence [6]. Garnet and Xider are open source databases from Microsoft® and YottaDB, respectively. The author of this article is sponsored by YottaDB and has contributed to its Lua API.

---

1  YottaDB and Redis are registered trademarks, respectively, of YottaDB LLC and Redis Ltd. All rights reserved.

2  For the sake of clarity, the code examples in this article omit certain error checks that are present in the article's code repository.

## 2 Materials and Methods

### 2.1 Lua language

The Lua language is admirably suited to implementing an API for key-value NoSQL [7] databases like Redis and YottaDB for several reasons. Firstly, like these databases, Lua is fast and light. The data structures of Lua (strings and tables) conveniently match the data structures of these databases. Secondly, Lua's classes allow operator overloading so that data can be accessed from Lua as if database values were native Lua types. Thirdly, Lua allows the same object-oriented API to be implemented using either Lua data structures or, if and when greater speed is required, using C data structures without changing the published API. The same API and benchmark may be implemented in other languages (as they have been in this article's code repository), but the focus in this article is on using Lua well.

Not only elegance, but speed, is important when implementing a Lua database API. For this reason, some of the efficiency-related API decisions are also discussed below, showing how Lua's object orientation has been used to enhance database speed.

### 2.2 Databases Distinctives

Redis has existed since 2009, becoming one of the most popular key-value databases, and is used in many web-enabled applications and as a high-speed query cache. Redis maintains high speed by keeping all data in memory and storing that memory image to disk using background processes. Like many databases, it is designed to convert all queries into a bitstream to transfer over a TCP/IP connection. Redis has two Lua interfaces: server-side scripting built into the database; and its Lua client – both of which have existed since Redis's earliest days. These two Lua options will add spice to our review.

YottaDB implements the M key-value database, originally conceived in 1966 and called MUMPS [8,9], and used extensively in medical, financial, and library systems today. The innovative feature and major strength of M is a tight coupling between its database and its native language (also called M), which YottaDB extends to other languages, including Lua. This close coupling enables YottaDB's very fast in-process database engine where processes use data structures in shared memory to cooperatively manage the database, avoiding slower forms of inter-process communication such as the conversion of queries into a bitstream for TCP/IP transfer.

To provide additional industry comparison points, two more databases have been benchmarked: Garnet [10] and Xider [11]. Both implement the Redis Serialisation Protocol (RESP), so they can automatically run the Redis version of the Lua benchmark. Garnet from Microsoft is designed to be extremely fast under concurrency conditions. Xider is a RESP-compatibility layer on top of YottaDB, designed to make the shared-memory concurrency speed of YottaDB available to Redis clients.

## 2.3 3n+1 sequences: a database benchmark

The Collatz Conjecture is one of the interesting – but unproven – results of number theory: stating that all $3n+1$ sequences are of finite length, ending at one. The sequence has numerous practical uses ranging from image encryption to software watermarking [1] – and now includes database benchmarking.

Each sequence is calculated starting with a number, $n$. To find the next number $n$ in the sequence: if the number is odd, multiply by three and add one (hence $3n+1$); otherwise divide by two. Then repeat until you reach one.

Significantly, most of these sequences 'meet up' with another sequence part way through (see Figure 1), so that the rest of the sequence only needs to be calculated once for both sequences. A program to calculate an entire block of these sequences (1 through a given integer $x$) can therefore benefit from looking up previous results, and by splitting the block of $x$ sequences into several blocks, the calculation can be split across multiple processors – where each process can share the others' work through, for example, a database.

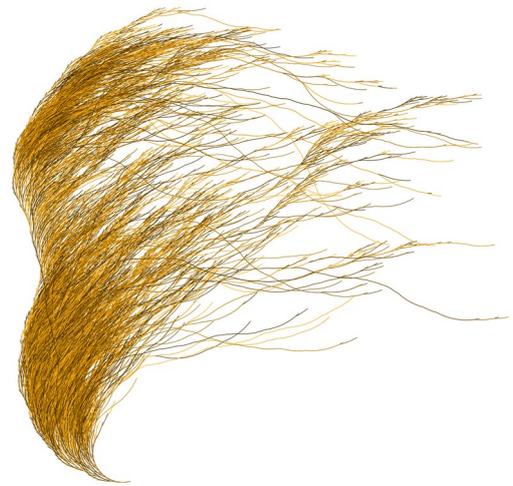

*Figure 1: 20,000 $3n+1$ sequences, visualised by K. J Runia's visualiser using Edmund Harris's visualisation rules showing how sequences meet up with one another [15]*

Such a program turns out to be highly appropriate as a database benchmark tool, as previously documented by Bhaskar [12]. He identified several factors that make it suitable. Firstly, it is a simple benchmark, for which he provided a reference implementation in the M language [2]. Secondly, it can be easily ported to a new language and API, making it easy to perform apples-to-apples comparisons between environments, as this article demonstrates. Thirdly, it performs a workload that is representative of real situations (such as web servers), where performance and throughput matter in concurrent environments. Fourthly, it can adjust its loading by simply altering the size of the sequence block to: alter runtime from seconds to hours, effectively exercise a range of machines from very small to large supercomputers, and can be set to utilise a database ranging from kilobytes to terabytes. Finally, its mathematical interest and legacy arouses curiosity, making it an attractive benchmark for programmers to implement.

To use this program as a benchmark for databases, we specify that the program must use parallel processes to gain speed, and those processes must use the database to communicate and share results from previously calculated sequences. The specific results required are four simple metrics: the length of the longest sequence, the largest number encountered in any sequence, and the total number of both database reads and writes required to perform the calculation. It will become significant later that database reads and writes used to perform process synchronisation should be excluded from these totals.

## 2.4 Machine specifications

The software and benchmarks were performed on a computer with an AMD Ryzen 7 5800X 8-core processor (each core exposing two execution contexts, which make it appear to have 16 CPUs), running at 2.2 GHz with 32 gigabytes of memory and a solid-state drive, operating under the SUSE Linux Enterprise Desktop 15 SP5 operating system.

# 3 Implementation

## 3.1 Redis server-side scripting language

Table 1 introduces the source code by means of a function that calculates a single $3n+1$ sequence into a database. This code runs as a server-side Redis script using Redis's built-in Lua interpreter.

*Table 1: redis3n1server.lua calculates a single $3n+1$ sequence into a database*

```lua
local function GET(key, default)  reads=reads+1  return redis.call('GET', key) or default  end
local function SET(key, value)  redis.call('SET', key, value)  updates=updates+1  end
local function HGET(key, field, default) reads=reads+1  return redis.call('HGET',key,field) or default  end
local function HSET(key, field, value, ...) updates=updates+1  redis.call('HSET',key,field,value,...)  end
local function sequence(n)
  local steps, highest, currpath = 0, 0, {}
  while  not HGET('step', n) and n>1  do
    currpath[steps] = n  -- log n as current number in sequence
    n = n%2>0 and 3*n+1 or n/2  -- compute the next number
    if n>highest then  highest=n  end
    steps = steps+1
  end
  if steps==0 then  return  end  -- if steps=0 we already have an answer for n
  if n>1 then   steps = steps+HGET('step', n)  end  -- add pre-existing path steps
  if steps > tonumber(GET('longest', 0)) then  SET('longest', steps)  end
  if highest > tonumber(GET('highest', 0)) then  SET('highest', highest)  end
  for i=0, #currpath do  HSET('step', currpath[i], steps-i)  end
end
```

An explanation of each section of code follows:

- The first four lines define wrapper functions that simplify counting of reads and writes to the database.

- Inside the 'sequence' function, the while loop calculates the sequence itself and stores it in local array *currpath*, exiting early whenever the current *n* is found in the database hash-table *step*.

- After the early exit case, the line beginning *if n > 1* simply adds the number of steps remaining in the rest of the sequence (stored in the database from a previous calculation).

- The two database values *longest* and *highest* are then updated with new results from this function. Note that this implementation would be subject to race conditions except that server-side scripting is blocking (solved for concurrency below, for both databases).

- Finally, the *for* loop stores into the database the number of sequence steps remaining at each step in the sequence – so that subsequent calls to *sequence()* can re-use the stored values.

There are several surprising characteristics of Redis server's Lua API:

1. It is curious that Redis server uses the syntax *redis.call('GET', …)* instead of defining more Lua-like functions such as *redis.get(…)* which the Redis client API provides (see below). For this reason our wrapper functions not only count reads and writes; they also shorten the long-winded *redis.call(func)* syntax.

2. Redis server does not permit the loading of user modules, so there is no way, for example, to share a module that provides your wrappers: wrappers must be implemented in every source file.

3. Most significantly, Redis server-side scripting is blocking, so no other process can execute at the same time as the function shown above. This makes server-side scripting impossible for concurrent use-cases like ours where we wish to gain speed from database parallelism.

This demonstrates that Redis server-side scripting is designed primarily for short tasks that need to be atomic (obviating the need for database transactions in certain cases), but it is not suited to the intensive parallel processing we require – primarily due to its blocking and limited language features.

No comparison with YottaDB server scripting is needed in this section, because YottaDB's server and client scripting are identical, and align more closely with Redis client scripting below.

## 3.2 Lua client scripts for both databases

### 3.2.1 Transactions

The problem of blocking Redis server code can be resolved using a Redis client. The core algorithm is identical, but database updates are no longer atomic since the server is no longer blocking. Therefore we must protect updates of *longest* and *highest* inside a database transaction. For example, suppose we calculate a new longest sequence of 100 entries, then we read and increase the value *longest* to 100, but before we can write it, another concurrent process updates *longest* to 200. Since we will then write our value of 100 to the database value *longest*, we will overwrite and lose track of the longer sequence of 200.

In order to update *longest* and *highest* inside a transaction, the two lines of code that update them above become more complex, as shown for both databases in Table 2.

*Table 2:* longest *and* highest *must now be updated inside a database transaction*

| Redis | YottaDB |
|---|---|
| ```repeat
  DB:watch('longest', 'highest')
    local db_longest = tonumber(GET('longest', 0))
    local db_highest = tonumber(GET('highest', 0))
  DB:multi()
    if steps > db_longest then
      SET('longest', steps)  end
    if highest > db_highest then
      SET('highest', highest)  end
until DB:exec()``` | ```ydb.transaction('batch', function()

    if steps > Longest:get(0) then
      Longest:set(steps)  end
    if highest > Highest:get(0) then
      Highest:set(highest)  end
end)()``` |

Table 2 shows how transactions are implemented in both Redis and YottaDB. Redis transactions require the user to specifically list variables to 'watch' against change during the transaction, and it requires the user to perform the transaction inside a loop that will repeat in the event that these 'watched' values were changed by another process.

The YottaDB API simplifies the transaction: it automatically identifies any variables that are adjusted during the transaction and performs its own repeat loop, if necessary. This automatic repeat is made possible by capturing the entire transaction in a single Lua function.

Storing the transaction code in a Lua function also provides future-proofing. A future YottaDB interface could transfer the entire transaction to the server end of a TCP connection for better performance and a lower chance of collision.

The other language difference is the option in the YottaDB API to access database nodes via Lua objects *Longest* and *Highest,* exemplified by the code `Longest:get(0)` which gets database node *longest* and defaults to 0. This object-oriented access is described in more detail below.

### 3.2.2 Object orientation

In order for the *GET()* and *SET()* functions to operate in Redis client like they did in Redis server, they need to be defined differently than our first Table 1 listing, to now use the *client* API as shown in Table 3.

*Table 3: wrapper functions for Redis client*

```lua
redis = require 'redis'
DB = redis.connect('127.0.0.1', 6379)
function GET(key, default)  reads=reads+1  return DB:get(key) or default  end
function SET(key, value)  DB:set(key, value)  updates=updates+1  end
function HGET(key, field, default)  reads=reads+1  return DB:hget(key, field) or default  end
function HSET(key, field, value, ...)  DB:hset(key, field, value, ...)  updates=updates+1  end
```

Redis uses a Lua object not to reference a database node, but to encapsulate a database connection, *DB* in Table 3, so that all subsequent database operations are performed on that object (using Lua's colon notation to access class methods). The wrapper functions for Redis client operation simply invoke class methods of the *DB* object. This is one possible use of object orientation.

By contrast, some database APIs implement an object-relational mapper (ORM) to make database access feel like native idioms within the language [13,14]. This is achieved using native language objects that invoke the database on demand to fetch data to a cache within the native language object, and store data back to the database using class methods. A simpler ORM may be implemented where the data is not cached, but is read directly from the database and written immediately to the database. If the database is an in-process and NoSQL database, such a minimal ORM can be more efficient without caching the data, provided that the database access itself is very fast.

YottaDB's Lua API permits an object-oriented system very similar to the ORM just described. The API allows objects to represent not only on a database connection but also database nodes. For example, a Lua object *Longest* can be defined to represent database node *longest* with the command `Longest = ydb.node('^longest')`. In our case, this helps us with our requirement to count database reads and writes. Instead of defining wrapper functions that count, as in the Redis example of Table 3, we simply override the *get* and *set* class methods of the node object as shown in Table 4.

*Table 4: Wrapper functions for YottaDB client*

```
ydb = require 'yottadb'
monitored_node, class, superclass = ydb.inherit(ydb.node)
reads, updates = 0, 0
function class:get(default)  reads=reads+1  return tonumber(superclass.get(self, default))  end
function class:set(value)  updates=updates+1  return superclass.set(self, value)  end
Step = monitored_node('^step')
Longest = monitored_node('^longest')
Highest = monitored_node('^highest')
```

Table 4 defines a subclass of *ydb.node* called *monitored_node*. This new class then has its *get* and *set* methods enhanced to count reads and writes. Any node object subsequently defined with the new class (e.g. *Step, Longest,* and *Highest*) will have its accesses counted.

You may have noticed that Table 4 only defines *get* and *set* methods whereas Table 3 also included hashing functions *HGET* and *HSET*. Separate hashing access functions are not needed in YottaDB since any node may be used as either a simple node, *X*, or as a multi-dimensional hash table: akin in some ways to Redis's single-dimensional hash table, *X[i]*, but also supporting multiple dimensions *X[i][j][…]*. So, in YottaDB, our *Step* variable is just an ordinary database node, but is accessed as a hash table using subscript *[i]*. This results in our previous while loop changing from Redis to YottaDB as shown in Table 5.

*Table 5: Access 'step' hash table via a function (in Redis) or an object (in YottaDB)*

| Redis | YottaDB |
|---|---|
| ```while  not HGET('step', n) and n>1  do``` | ```while  not Step[n].__ and n>1  do``` |
| ```  ...``` | ```  ...``` |
| ```end``` | ```end``` |
| ```...``` | ```...``` |
| ```if n>1 then  steps = steps+HGET('step', n)  end``` | ```if n>1 then  steps = steps + Step[n].__  end``` |

In Redis you must access the *step* database value using a function. With the YottaDB API you can access it directly via the Lua object *Step* and using Lua's standard indexing symbols '[]' – but in this case accessing the database node rather than a Lua table. In this case, the value of *Step[n]* is accessed via its '__' value property, which is the API's shorthand for its *get* method: `Step[n]:get()`. The '__' property may also be used to assign values to the node as shorthand for *set*.

Table 6 adds a function to calculate not just one sequence, but a block of sequences.

*Table 6: Fetch the next available block of sequences from a database table and calculate each one.*

**Redis**

```lua
function nextblock()
  index = 1
  while DB:hget('blocks', index+1) do
    if DB:hincrby('blocks',index..'-taken',1)==1 then
      local first = DB:hget('blocks', index)+1
      local last = DB:hget('blocks', index+1)
      for n=first,last do  sequence(n)  end
    end
    index = index+1
  end
end
```

**YottaDB**

```lua
Blocks = ydb.node('^blocks')
function nextblock()
  index = 1
  while Blocks[index+1].__ do
    if Blocks[index].taken:incr() == '1'  then
      local first = Blocks[index].__+1
      local last = Blocks[index+1].__
      for n=first,last do  sequence(n)  end
    end
    index = index+1
  end
end
```

Multiple processes will be set up (below) to calculate different blocks concurrently. To manage these processes, they share a database array called *blocks,* of sequences to be calculated (e.g. *0, 10, 20, 30, 40* – where the first block of numbers to calculate is 1-10, the second is 11-20, etc.). From this list each process fetches the next available block of sequences to calculate, and marks that block as being calculated.

Semantic differences are as follows:

- Redis code must not use our *HGET* function (defined in Table 1) to access *blocks* because HGET counts database reads – which may only be counted for sequence calculations, not for process management. By contrast, YottaDB code accesses the database using standard object-oriented idioms, and simply defines *Blocks* in the first line with *ydb.node()* rather than *monitored_node()*, ensuring its reads will not be counted.

- Redis's code, as highlighted, marks a block as 'taken' by appending '-taken' to the hash table key name. The YottaDB code takes advantage of multi-dimensions subscripts to achieve the same goal in a slightly more elegant way using a subscript called 'taken'.

### 3.2.3 Process management and busyloops

Properly benchmarking a database needs concurrent processes accessing it. In order to do this we add process management logic around our $3n+1$ sequence calculator, using database semaphores, somewhat like traffic lights, to communicate when processes should start and finish.

This process management logic is demonstrated in the *manage_workers* function in Table 7. The function's first parameter (*nWorkers*) is the number of worker processes to start. The second parameter is the Lua array of *blocks* to be calculated, that we will store in the database to pass to each new process.

Table 7: Process management

| Redis | YottaDB |
|---|---|
| ```lua
function manage_workers(nWorkers, blocks)
  for key, value in pairs(blocks) do
    HSET('blocks', key, value)
  end
  SET('trigger', 0)
  SET('queued', nWorkers)
  DB:del('worker')
  for worker=1, nWorkers do
    fork_func('subprocess', worker)
  end
  -- wait for all jobs to start
  while GET('queued') ~= '0' do  sleep(0.1)  end
  SET('trigger', 1)
  local start = time()
  while DB:hlen('worker') ~= 0 do sleep(0.1) end
  return time() - start
end
function subprocess()
  local pid = getpid()
  DB:hset('worker', pid, 1)
  DB:incrby('queued',-1)  -- flag ready to start
  while DB:get('trigger') ~= '1' do sleep(0.1) end
  nextblock()
  DB:incrby('reads', reads)
  DB:incrby('updates', updates)
  DB:hdel('worker', pid)  -- flag finished
end
``` | ```lua
function manage_workers(nWorkers, blocks)

  Blocks:settree(blocks)

  Trigger:grab()
  Queued.__ = nWorkers
  -- YottaDB auto-releases dead process semaphores
  for worker=1, nWorkers do
    fork_func('subprocess', worker)
  end
  -- wait for all jobs to start
  while Queued.__ ~= '0' do  sleep(0.1)  end
  Trigger:release()
  local start = time()
  Finished:grab() Finished:release()
  return time() - start
end
function subprocess()
  local pid = getpid()
  Finished[pid]:grab()
  Queued:incr(-1)  -- flag ready to start
  Trigger[pid]:grab()  Trigger[pid]:release()
  nextblock()
  Reads:incr(reads)
  Updates:incr(updates)
  Finished[pid]:release()  -- flag finished
end
``` |

The first Redis *for* loop simply stores the Lua array *blocks* into a database array *blocks*. The YottaDB API *settree* method does the same. The code then creates worker processes that each run the *subprocess* function (which runs nextblock to grab the next available block of sequences, compute them, and repeat until all blocks are computed).[3] Finally, it waits for all worker threads to complete.

The key differences between APIs are highlighted. All of these have to do with the availability of locks in YottaDB, as follows:

- The Redis code must use ordinary database nodes to synchronise database operations, but YottaDB provides semaphores (referred to as *locks* in its documentation) – which have the advantage that they are automatically released when a process terminates.

- For accurate timing, the concurrent *subprocess* functions all start calculating at the same time. This is coordinated by the database node *trigger*. In YottaDB, this occurs because the parent process grabs Trigger. YottaDB locks (i.e. semaphores) are hierarchical in the sense

---

3  The *fork_func* function is an OS-specific function to create a subprocess that runs the given Lua function. Its source is not shown here but may be found in the source code links provided [2].

that any other process wanting to access a subscript of a grabbed lock must wait until the parent lock is released. Since the *subprocess* function grabs a subscript of Trigger before starting calculation, the process sleeps until *manage_workers* releases *Trigger*.

- The Redis timing solution is based on an ordinary database node with the drawback that a polling busyloop is required in each process to test for the *trigger* flag.

- The parent process waits for all subprocesses to complete using the same technique in reverse. In YottaDB, each subprocess grabs a subscript of the *Finished* lock. Once the parent determines that all subprocesses have started, it also tries to grab the *Finished* lock – which puts it to sleep until all subscripts of that lock have been released. The parent process in the Redis solution must once again busyloop for the entire calculation, polling until all processes have finished.

In terms of semantics, the use of Lua index notation (e.g. *Finished[pid]*) is an elegant way for the API to provide access to hierarchical lock objects, just as it was used previously to access multi-dimensional database objects.

# 4 API Object Internals

## 4.1 Maintaining API speed using Lua objects

API speed is as important as elegance when implementing a database API, especially when a database maintains speed as a key feature (see the benchmark results below). Here we dig below the surface to unearth some of the decisions involved in developing the Lua API, to maintain speed in conjunction with Lua elegance. It will be necessary to focus on the YottaDB API alone because the Redis API does not use Lua objects to represent database nodes, which is what makes this technique possible.

Central to the ability to maintain API speed is the design choice to represent the path to a database node as a Lua object. This lets the API cache its working data in the object, which in turn enables multiple actions to be performed on that database node without having to regenerate that data.

For example, each node operation in YottaDB requires addressing the node by means of a 'path' of Lua strings. That node-path must be passed as parameters to each database call as, for example, a Lua table of strings. For a slightly contrived example:

*Table 8: node access **without** node-path stored in the node object*

```
birthday = ydb.get('demographics', {'country', 'person', 21, 'birthday'})
age      = ydb.get('demographics', {'country', 'person', 21, 'age'})
if  date >= birthday and age != nil  then
    ydb.incr('demographics', {'country', 'person', 21, 'age'})
end
```

Instead, the API lets the node object itself contain the re-usable node-path address:

*Table 9: node access **with** node-path stored in the node object*

```
person = ydb.node('demographics').country.person[21]
if  date >= person.birthday:get() and person.age:get() != nil   then
    person.age:incr()
end
```

Notice that the path to the database node is now stated only once and stored in the *person* object. Now that there is a *person* object to hold the node path, it can be cached within the object in the format required by YottaDB shown below, speeding up successive commands to the database.

## 4.2 Object metadata and the Lua-C interface

### 4.2.1 Iteration 1

We saw above how, in principle, Lua's ability to store metadata in objects can speed up database access. Table 10 shows a first attempt to represent the object's internal data structures:

*Table 10: node object's internal data structure*

```
person = {
    varname = "demographics"
    subsscripts = {"country", "person", "21", "gender"}
    cachearray = <userdata>
}
```

In this example, the Lua strings (subscripts) that form the node path and that were passed into *ydb.node()* at object creation, are referenced by object fields *varname* and *subscripts*. The last field, *cachearray* is of a Lua type called *userdata* which permits storage of C data. In this case, the C data is a C array pointing to these same Lua strings, which will be passed to YottaDB's C API for fast access to the database node. Referencing and then pointing to Lua strings like this is possible since Lua strings are immutable, and remain fixed in memory (until they are no longer referenced). The purpose of this mechanism is to prevent having to make copies of the Lua strings in C memory which would take time and space, considering that subscripts may be up to a megabyte in size.

The API cannot be pure Lua, but must have a C portion: firstly, to call the YottaDB C API functions. Secondly, the *userdata* type can only be created by a C function that calls the Lua API function *lua_newuserdatauv(datasize)* and then populates it with the C array of strings.

The metadata structure shown above was tested and worked fine, except that node creation ended up being unexpectedly slow. Profiling showed that creation of a Lua table is slow compared to a C array. Each table requires several steps: memory allocation (*malloc*), creation of the hash table, the numeric index of the Lua table, and linkage into the Lua garbage collector. Furthermore, the structure shown above requires two Lua tables: one for the *subscripts* and one for the *person* object itself (because a Lua object is typically a table). Finally, the new object also required creation of the *userdata* type, which has the same overhead as a Lua table, except without the hash table.

### 4.2.2 Iteration 2

The beauty of using a Lua object for the API is that Lua provides more than one mechanism for implementing an object. This means that a faster version of the API can be created later using a different mechanism, without changing the published API at all. In this case, the second iteration of the API uses the Lua feature that a *userdata* type itself can implement the object, *instead of* a Lua table. This reduces object creation overhead by eliminating creation of the two Lua tables shown above.[4]

Furthermore, given that the use of megabyte-sized node-path subscripts is very rare, and copying strings is incredibly fast, we may consider relaxing the previous constraint that avoided copying Lua strings. Considering that memory allocation takes time, and all string lengths are known in advance when *ydb.node()* is invoked, it is possible to pre-allocate space for all the strings in a single memory allocation using *lua_newuserdatauv(datasize)*. The concern over using duplicate string memory is also negligible: although twice the memory size of the strings will be required very briefly while they are copied, the original strings, if no longer referenced are free for garbage collection immediately after node creation.

The C component of the API is beyond the scope of this article. However, for conceptual purposes it is sufficient to present a C structure suitable for the final *userdata* type:

---

[4] It is possible to use either *full userdata* or *light userdata* types to represent a Lua object. In this case, Lua's *full userdata* type is used because, unlike *light userdata,* it will automatically be garbage collected when the object is no longer referenced.

*Table 11: node object's internal structure in C as a Lua userdata*

```
typedef struct node {
  int stringcount;              // number of strings in the string pointer array
  int stringdata_size;          // size allocated for all strings in stringdata
  ydb_buffer_t stringarray[];   // YottaDB-required array of pointers to strings' data
  char stringdata[];
} node;
```

This C table stores the same essential information as the structure in Table 10. The field *stringdata* contains both the *varname* and *subscripts* strings. The *stringarray* substitutes for *cachearray* of Table 10 which stores the C array required by calls to YottaDB. The lengths of *stringarray* and *stringdata* are stored in the respective fields *stringdata_size* and *stringcount* (these lengths were inherent in the Lua types of Table 10 but must be explicit in C).

With this internal data structure as a C structure in a Lua *userdata* type, the API implementation no longer needs to create any Lua tables when creating a new node object, making the second implementation considerably faster.

## 5 Benchmarks and Results

### 5.1 Results

This section compares the speed of the two databases using the $3n+1$ program that we developed above. YottaDB and Redis are presented as well as Xider and Garnet as industry comparison points, outlined in 2.2 . The results are tabulated in Table 12 and graphed in Figure 2.

*Table 12: Time in seconds to calculate 100,000 sequences*

| Database system | 32 processes | 1 process |
|---|---|---|
| YottaDB | 0.4 | 1.2 |
| Redis | 5.0 | 12.9 |
| Garnet | 2.5 | 15.8 |
| Xider | 3.1 | 16.4 |

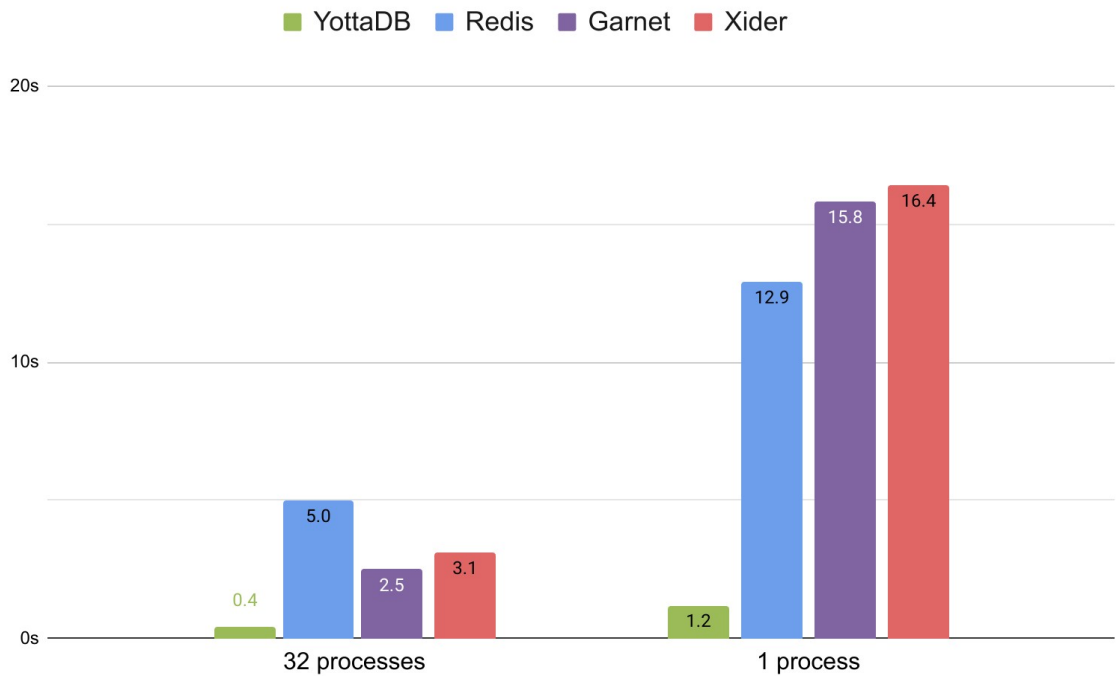

*Figure 2: Time to calculate 100,000 sequences*

Figure 2 presents the time required to calculate 100,000 3n + 1 sequences. Each database test had full and exclusive access to all CPUs in the machine.

## 5.2 Discussion

Two sets of results were taken. The right-hand set shows the 3n + 1 algorithm running in just one process, which does not use concurrent access to the database, and so does not fully utilise the machine. By contrast, The left-hand scenario runs on twice as many processes as the number of CPUs in the machine, ensuring that we fully exercise not only the machine but also the operating system's preemptive multitasking system, the 3n + 1 calculation database client, and the database's transaction system.

Database server and client were both running on the same machine, so there is no physical network traffic to slow down the software for any of the database choices. Nevertheless, YottaDB (the green bar) was found to be an order of magnitude faster than the other databases. However, this benchmark result should not be taken to imply that YottaDB's Lua client alone is faster than the Redis Lua client. Instead, this is partly to be explained as an outcome of YottaDB's unique method of in-process data sharing, which is able to access database memory directly and does not need to turn database queries into a bitstream for TCP/IP transfer.

Xider uses the YottaDB database back-end, but provides a RESP-compatible layer on top of YottaDB. In this case, both Xider and Redis perform data serialisation which allows them to operate over a TCP/IP network, providing a more apples-to-apples comparison. In the case of the multi-process scenario, Xider is seen to be faster than Redis, but this time for a different reason: it is not merely a single Redis's server that manages concurrent access. Rather, it has clients connecting to *multiple* RESP server processes, thus running RESP parsing in parallel at both ends. Each server connects then to the one database via YottaDB's in-process shared memory mechanism.

Ultimately, benchmarking databases provides metrics that can reduce costs. For example, the common use of key-value databases to cache queries can be fulfilled by any of the benchmarked databases, yet this benchmark shows a 10-fold difference in performance between them. In this scenario, a faster database could significantly reduce the cost and number of caching machines required.

## 6 Conclusion

We described the implementation of a novel $3n+1$ benchmark tool for concurrent database access using the Lua language features of two different database systems: Redis and YottaDB. The core benchmark algorithm was initially presented in Redis's Lua server scripting language. Although server scripting is fast, it was found to have limited Lua language features and no concurrency, making it unsuitable for the full requirements of this benchmark. Consequently, Lua language features were then compared in a Redis client and a YottaDB client.

We discussed Lua language differences in four areas. Firstly, the Lua implementation in YottaDB offers a way to encapsulate transactions in a Lua function, removing the need for the programmer to specify variables for change tracking, and minimising concurrency conflicts during transaction processing. Secondly, for these NoSQL databases, a more complex ORM layer can be avoided by implementing simple object-oriented access to nodes, providing database syntax that feels natural in Lua, with class methods that can be powerfully overridden. Thirdly, we demonstrated the use of standard Lua indexing syntax for accessing hierarchical database process semaphores to co-ordinate process interaction. Fourthly we showed how Lua's object orientation can be used to cache working data to speed up database node access using data representations in either Lua or C, depending on speed requirements.

Finally, we showed how the $3n+1$ benchmark tool can be used to compare the speed of different databases, and determine the performance benefit of unique database architectures such Redis's

in-memory database and YottaDB's in-process data sharing. Any such speed benefit can in turn mean that the database can run at higher efficiency under heavy loads.

In summary, the proposed 3n + 1 benchmark tool was shown to be ideal not only to exercise the speed of a given database but also to compare its language features.

## 7 Abbreviations

**API**: Application Programmer Interface

**CPU**: Central Processing Unit

**ORM**: Object-Relational Mapper

**OS**: Operating System

**SQL**: Structured Query Language

**RESP**: Redis Serialisation Protocol

**TCP/IP**: Transmission Control Protocol operating over an Internet Protocol network

## 8 CRediT author statement

**K. S. Bhaskar**: conceptualisation; software – original M-language version; writing – review;

**Berwyn Hoyt**: conceptualisation; methodology; software; validation; analysis; investigation; writing; visualisation.

## 9 Acknowledgments

Thanks to K. S. Bhaskar for proofing, proposing 3n + 1 as a benchmark tool and providing the original M-language reference implementation. Thanks also to YottaDB for sponsoring this article.